# Determination of Handedness in a Single Chiral Nanocrystal via Circularly Polarized Luminescence

Eitam Vinegrad[1,†], Uri Hananel[2,†], Gil Markovich[2] and Ori Cheshnovsky[2,*]

1. School of Physics, Raymond and Beverly Faculty of Exact Sciences, Tel Aviv University, 69978 Tel Aviv, Israel.
2. School of Chemistry, Raymond and Beverly Faculty of Exact Sciences, Tel Aviv University, 69978 Tel Aviv, Israel.

[†] These authors contributed equally to this work

*Corresponding author: orich@post.tau.ac.il

August 16, 2018

**The occurrence of biological homochirality is attributed to symmetry breaking mechanisms which are still debatable[1]. Studies of symmetry breaking require tools for monitoring the population ratios of individual chiral nano-objects, such as molecules, polymers or nanocrystals. Moreover, mapping their spatial distributions may elucidate on their symmetry breaking mechanism. Recently, researchers have utilized differential scattering or circular dichroism microscopy to identify chirality on individual plasmonic nanostructures and inorganic nanocrystals[2–7]. However, these measurements are prone to optical system artifacts. While luminescence is preferred for detecting single particle chirality, the typical low optical activity of chromophores limits its applicability[8,9]. Here, we report on handedness determination of single chiral lanthanide based nanocrystals, using circularly polarized luminescence, with a total photon count of $2\times10^4$. We also utilize a machine learning approach[10] for correlative**

**microscopy to determine and spatially map handedness of individual nanocrystals with high accuracy. This technique may become invaluable in studies of symmetry breaking in chiral materials.**

Circularly polarized luminescence (CPL) spectroscopy is based on the unequal emission of left- and right- circularly polarized light by a chiral species[11]. The magnitude of this difference depends on the properties, as well as on the degree of enantiopurity, of the examined material. The measured quantity is the emission circular intensity differential (ECID), $I_L - I_R$, where $I_L$ and $I_R$ are the intensities of the emitted left- and right-handed circular polarizations, respectively. ECID can be used to monitor the handedness of a light emitting sample, and has been measured for chiral organic molecules[12], and semiconductor nanocrystals (NCs) coated with chiral molecules[13,14]. The strongest ECID was measured in chiral lanthanide complexes[15,16]. Recently, large ECID was measured in inorganic lanthanide phosphate NCs which crystallize in a chiral space-group[17]. These experiments led us to strongly suspect that at certain synthesis conditions only one NC enantiomer is formed. However, until recently there has been no experimental method which could quantify the enantiomer population ratios in a chiral NC sample within a reasonable amount of time and confidence. Here we describe a methodology which allows us to perform high-throughput single particle quantitative assessment of the enantiomer population ratios of lanthanide based chiral NCs.

Rod-shaped $Eu^{3+}$ doped (5%) $TbPO_4 \cdot D_2O$[18] lanthanide NCs were synthesized by a simple aqueous procedure described in the Methods section. The lanthanide NC handedness was directed with either D- or L- tartaric acid present in the synthesis solution, and we label the resulting crystals as 'D' and 'L' accordingly. The NCs had typical dimensions of 1.4 µm × 50 nm.

Emission measurements of NCs dispersed in D$_2$O exhibited sharp peaks typical of Eu$^{3+}$ in the visible range, attributed to the $^0D_5 \rightarrow \ ^7F_J$ transitions[19] where $J = 1,\ldots,4$. In both ensemble and single particle measurements we excite the Tb$^{3+}$ ions (at 365 nm and 488 nm, respectively). The Tb$^{3+}$ ions transfer the excitation energy to the Eu$^{3+}$ ions, which emit after a characteristic lifetime on the order of a millisecond[20,21]. The NCs displayed ECID lines with varying relative intensities throughout the emission spectrum, with inverted polarity for the two opposite enantiomers (Fig. 3d), as expected from a chiral emitter. The degree of dissymmetry is given by the dimensionless dissymmetry factor $2\frac{I_L-I_R}{I_L+I_R}$, taking values between $\pm 2$. High dissymmetry emission lines display the greatest relative line-shape difference between the two enantiomers, and can thus serve for identification of the handedness of single particles. The 596 nm and 650 nm emission lines displayed particularly high dissymmetry values (0.15 and ≥0.4, respectively)[17]. The majority of the emission at and above 590 nm stems from the Eu$^{3+}$ dopant ions in the interior of the NC, as surface emission is quenched[20,22]. Since the NCs are all single crystals of a particular handedness[17], all of the emitting ions should be located in the same type of local chiral environment and thus produce the same ECID polarity. Taking into account the average diameter of the nanorods and the size of the diffraction-limited excitation spot we estimate the numbers of emitters illuminated by the excitation beam to be in the order of ~10$^5$. The low absorption cross section of these rods together with a low quantum efficiency (~30%) leads to ~200 photons/s at our EM-CCD detector for the 596 nm peak. Considering the overall optical transmission and the collection solid angle, this translates to a total of ~8000 photons/s emitted by this particle at the particular emission line, which is several orders of magnitude lower than a typical single fluorescent molecule like Rhodamine 6G.

For single particle ECID measurements, the NCs were drop-cast on a Formvar film coated copper grid to enable both transmission electron microscopy (TEM) and luminescence microscopy for the same population of individual NCs. Single particle distribution was mapped on the grid by TEM. The grid was then placed on a low fluorescence fused-silica cover glass, and the same particles were measured by CPL microscopy, using the setup illustrated in Fig. 1.

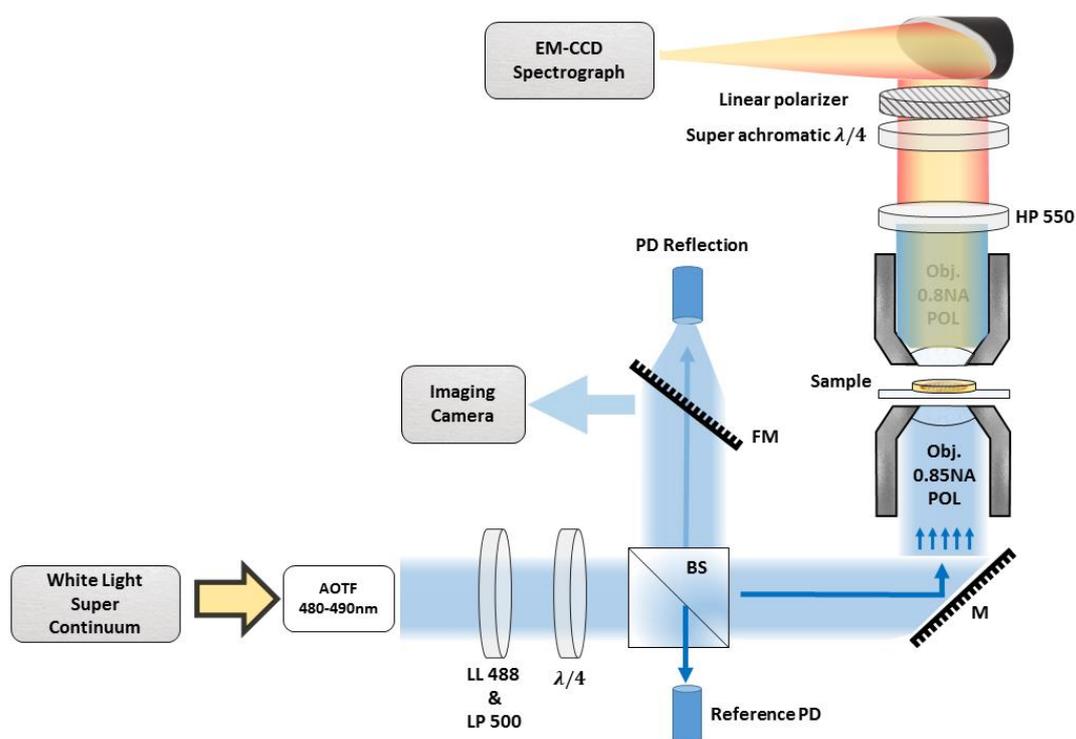

*Fig. 1 – Schematic of the experimental setup. The broadband supercontinuum source is filtered by an AOTF to select the 488 nm excitation line. The excitation laser is further cleaned via 488 nm laser line (LL) and 500 nm low pass (LP) filters. A λ/4 waveplate is used to convert the excitation polarization into circular, to remove the dependence of the NCs orientation on excitation polarization. Then, the excitation beam is passed through 50:50 non-polarizing beamsplitter, separating the light to a reference beam and a probe beam. The probe beam is focused on the sample, which is scanned by a piezoelectric motor. This enables the detection and localization of single NCs. The transmitted light is collected by another objective and is blocked by a 550 nm high pass filter (HP) to transmit only the NCs luminescence. The emitted light is then passed through a super achromatic λ/4 waveplate followed by a linear*

*polarizer at 45° to the fast axis of the λ/4 waveplate to a photodiode (PD) detector. This enables measurement of circularly polarized emission of a specific handedness. ECID is measured by subtracting this spectrum from another in which the polarizer is rotated by 90°. The reflected light is used to locate the NCs using extinction (PD), while the camera is used to correlate the locations of the particles to the TEM micrographs.*

These nanorods exhibit a large variety of peak intensities. This phenomenon is depicted in Fig. 2a, showing several unpolarized emission spectra from a solution ensemble, an aggregate of many NCs as well as two single NCs. While the general line-shapes and peak wavelengths of these spectra are similar, they exhibit distinct variance in the relative peak intensities, which most likely stems from different defects and doping efficiency in each NC[23,24]

Since the ECID signal is usually a small fraction of the luminescence signal, it is desirable to employ fast modulation between the right (r-CPL) and left (l-CPL) handed circular polarization luminescence, combined with lock-in detection. The difficulty in such approach is that polarization modulation of the emitted light (e.g. using liquid crystal variable retarder or photoelastic modulator) requires wavelength dependent tuning and temperature sensitive retardation calibration. Hence, in a wide wavelength range, the emitted light is collected sequentially by scanning over the relevant range (in this case 580 nm - 720 nm). The large dissymmetry values of several emission lines enabled us to use two other schemes for identifying the handedness of the individually measured particles:

I. Measuring ECID by subtracting two accumulated spectra of r-CPL and l-CPL. II. Analyzing the spectral line-shape of only r-CPL or l-CPL to deduce the handedness.

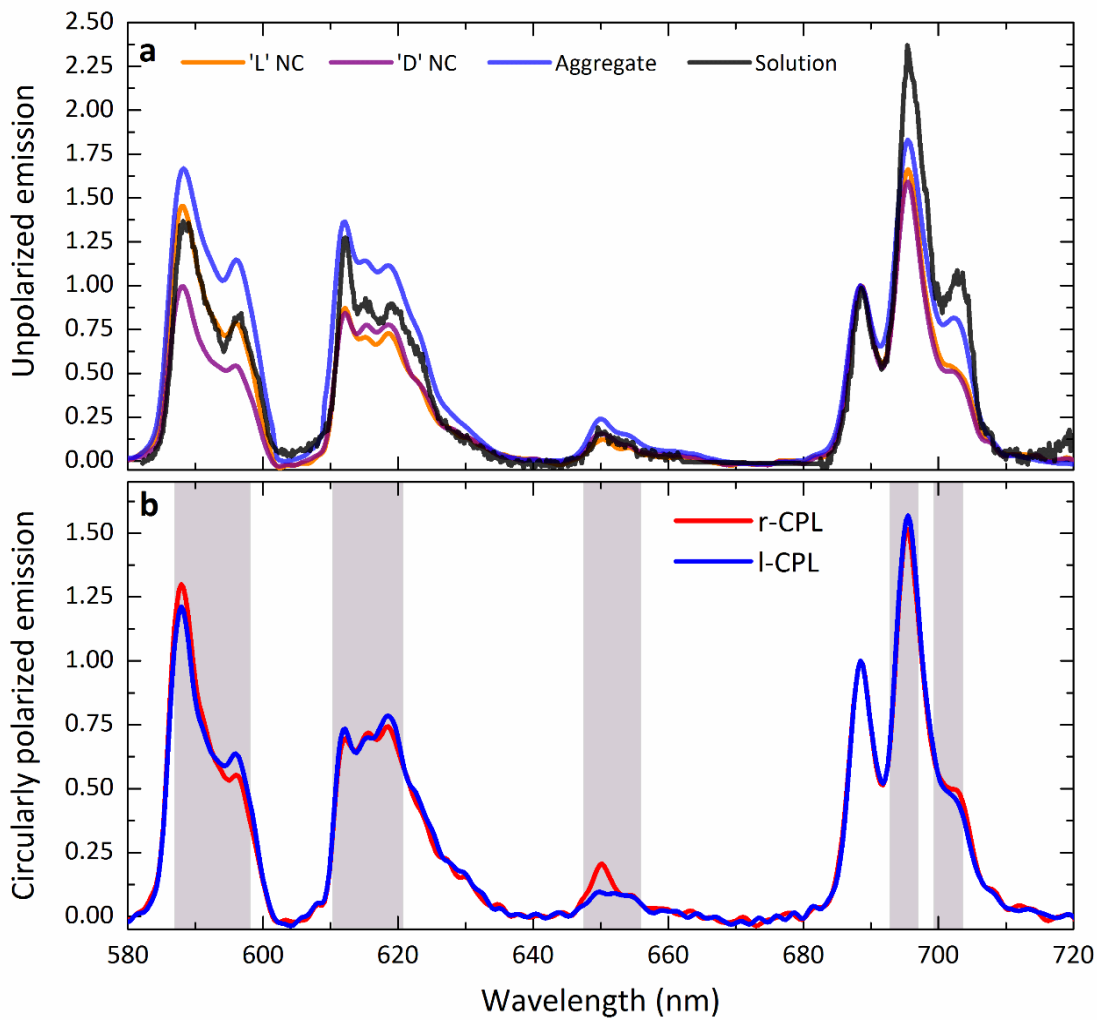

*Fig. 2 – **a**, Unpolarized luminescence spectra from solution ensemble (black), an aggregate of many NCs (Blue) as well as a typical 'L' (orange) and 'D' (purple) lanthanide NCs. **b**, right- (red) and left- (blue) handed circularly polarized luminescence (r-CPL and l-CPL, respectively) from a single 'D' lanthanide NC. Shaded areas mark the wavelength ranges that exhibit significant CPL signal. All left circularly polarized spectra were normalized by scaling with respect to the same particle right circularly polarized spectra to compensate for the polarization dependent light collection efficiency of the two orthogonal circular polarizations. Scaling factors were extracted from the points which displayed the lowest dissymmetry in ensemble measurements.*

To measure the entire r-CPL or l-CPL spectrum in a single measurement cycle we used a super achromatic λ/4 waveplate combined with a linear polarizer, switching the polarizer between + and −45° with respect to fast axis of the waveplate, respectively.

Fig. 2b shows the r-CPL and l-CPL from a single NC. These spectra show areas with significant emission dissymmetry (>$10^{-2}$, shaded in grey).

Fig. 3a,b shows the ECID spectra ($I_L - I_R$) of several single NCs of both handedness. These spectra show that some of the ECID peaks are stable between different NCs, e.g. the 596 nm and 650 nm peaks, while other peaks, like the 612 nm peak, are particle dependent and vary significantly between individual particles. These spectra, which clearly define the handedness of these individual particle, were obtained with a total photon count of $2\times10^4$ photons (collected during 100 s) originating from about $10^5$ $Eu^{3+}$ ions located within the diffraction limited illuminated volume of an individual NC. Fig. 3c shows the average ECID over seven particles for each of the two enantiomers, which compares quite well to the ensemble average measurement in a colloidal solution shown in Fig. 3d.

The measurement of ECID over a large distribution of individual particles more than doubles the total measurement time compared to only measuring r- or l-CPL, and increases the experiment's sensitivity to sample drifts, limiting the usefulness of this method. Looking back at the raw circularly polarized emission spectra in Fig. 2b, we note that the high dissymmetry of some of the emission peaks leads to slight yet discernible differences in the luminescence spectral line-shape between the two orthogonal circular polarizations.

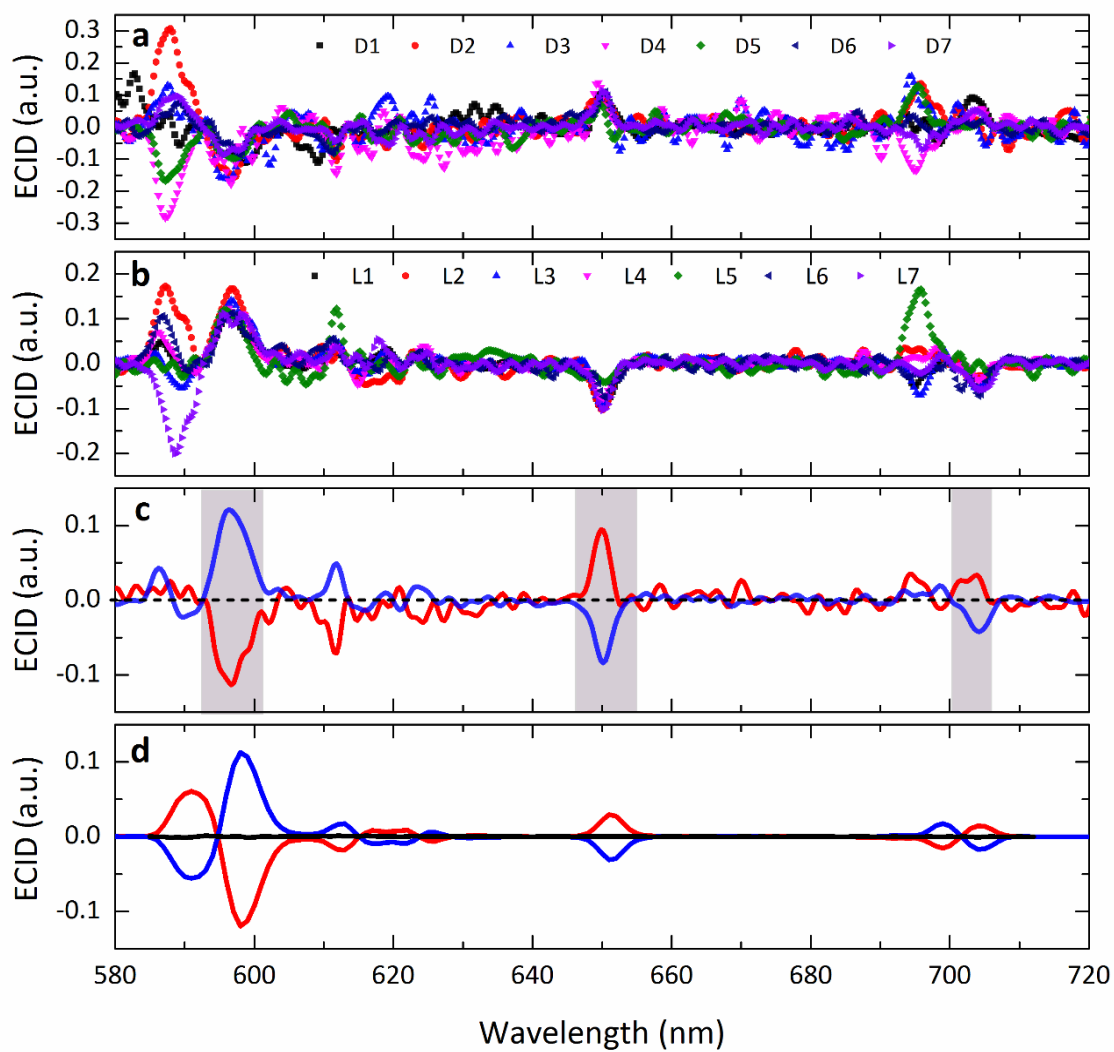

*Fig. 3 – **a**, ECID spectra of seven different single 'D' lanthanide NCs. **b**, ECID spectra of seven different single 'L' lanthanide NCs. The spectra in (**a,b**) were normalized by the 688 nm peak intensity, and corrected for the polarization collection efficiency before subtraction of $I_L - I_R$. **c**, Average ECID spectra for the seven lanthanide 'D' NCs (red line), and seven 'L' NCs (blue line). Shaded gray areas mark spectral regions for which the ECID of the single NCs has the lowest variance between the measured NCs. These spectral regions were then used for the input features for the machine learning classifier. **d**, Ensemble ECID spectra of 'L' (blue), 'D' (red) and the racemic (black) solution samples. These ensemble spectra are represented quite well by the few particle average spectra shown in **c**. The spectra in **d** was scaled by a constant to be of the same range as in **c**.*

These line-shape differences, which are inverted between the two NC enantiomers, allowed us to determine the NC handedness without measuring ECID spectra, i.e., by only measuring only a single circular polarization. This single measurement also minimizes the offset of ECID values originating from residual asymmetry of the experimental setup response to right- and left-handed circularly polarized light.

In order to establish the use of line-shape analysis of single r- or l-CPL spectra to determine the NC handedness, we measured many single particles' r-CPL spectra over two different samples which, according to ensemble solution ECID measurements, were expected to be pure 'L'- and pure 'D'-NCs[17]. For handedness determination based on spectral features we employ a machine learning Support Vector Machine (SVM) classifier. For the SVM training data we measured 64 individual NCs from the 'D' sample and 60 individual NCs from the 'L' sample.

The SVM algorithm input consisted of four distinct features of the luminescence line-shape, and the corresponding NC handedness. Since the luminescence intensity is not a quantitative attribute, as it is dependent on NC size, individual doping efficiency and illumination intensity, we normalized each peak of interest to a neighboring peak which exhibits a lower dissymmetry factor in ensemble measurements. Fig. 4a shows a scatter plot of the training data using three of the four selected spectral features. The scatter plot displays a clear separation surface between the two NC enantiomers. Fig. 4b outlines the four normalized spectral features, shows their individual scatter plots and the classification accuracy of the SVM classifier used with only that single feature. Using the four spectral features, the SVM algorithm classified the training data with an error rate of less than one percent (0.8%). Only one NC from the 'L' sample was classified as a 'D' NC. A close examination of this NC line-shape in comparison to the line-shape of other 'L' and 'D' particles lead us to the conclusion that this specific particle is indeed a 'D' handed NC, essentially bringing the classification accuracy to 100%. The discovery of a 'D' NC impurity in the supposedly enantiopure 'L' NCs sample further demonstrates the importance of the single particles handedness study, and the robustness of the SVM classifier.

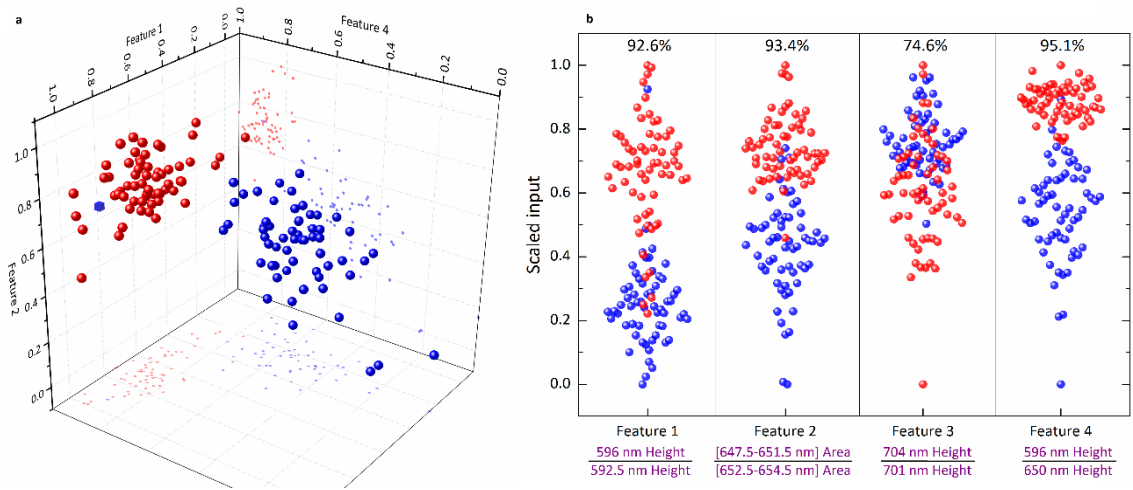

*Fig. 4 – **a**, Scatter plot of the classification training data from presumably 'L' and 'D' enantiopure solutions. Accordingly, blue and red spheres are the correctly classified NC out of the training set. The blue cube is a NC classified in contradiction to its expected handedness. Data points were rescaled to the range [0, 1] for each axis independently. The scatter plot uses three out of the four features that led to the highest classification accuracy. Feature 1 is the ratio of the 596 nm peak to that of the 592 nm shoulder. Feature 2 is the area ratio between the 650 nm peak and the 653 nm shoulder. Feature 4 is the ratio of the 596 nm peak to that of the 650 nm peak height. **b**, Scatter plots for each of the individual data features used for data training and classification. The percentage over each feature represent the SVM training accuracy reached when using only that feature for training.*

Once the SVM classifier was trained, we moved on to a racemic NC sample; an equal mixture of the 'D' and 'L' NCs, which displayed zero CPL in ensemble measurements. We were able to predict the handedness of individual NCs in this sample. These prediction were randomly checked over 7 particles of each handedness, with 100% success (shown in Fig. 3a,b). For the 162 NCs measured out of the racemic mixture we found that the D to L particle ratio is 90:72, so that the probability of finding a 'D' enantiomer is $P(D) = 0.555 \pm 0.064$, which is within the statistical error margin from the expected 50% ratio. Fig. 5 shows TEM micrographs of the racemic NCs sample with individual NCs artificially colored according to the SVM classifier handedness analysis.

Unexpectedly, the spatial distribution of the two opposite NC enantiomers in the racemic sample seems to be heterogeneous: NCs of the same handedness are more likely to be found in proximity to their own kind. This effect, which has been observed for other salts[25], may be related to symmetry breaking and chiral amplification recently observed in the synthesis of such NCs[17], and will be further studied in the future using the technique presented here.

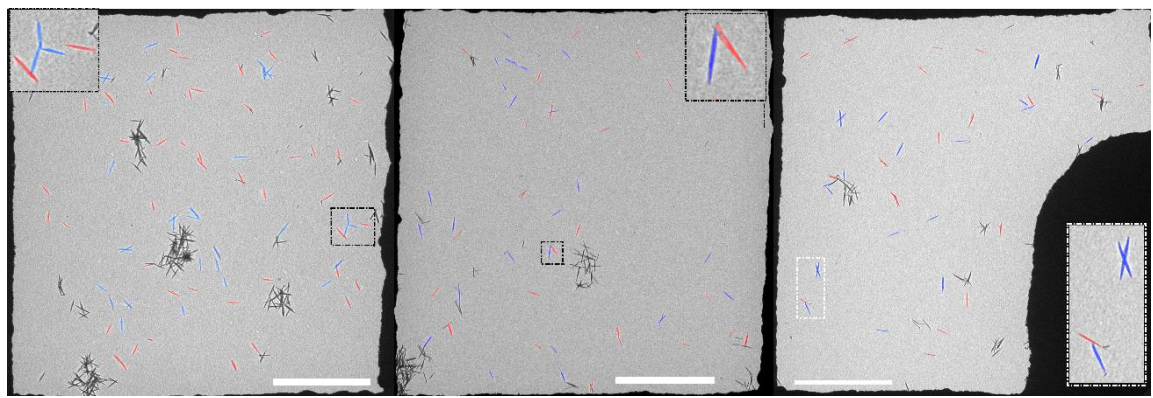

*Fig. 5 – TEM micrographs of lanthanide NCs of a racemic solution. The sample was prepared by drop-casting a NC racemate solution on a Formvar film. The measured NCs were color coded according to their classification results, blue for the 'L' enantiomer and red for the 'D' one. Scale bars are 10 μm. The un-colored NCs were not measured. Insets are magnifications of the areas surrounded by the dashed rectangles, showing the high spatial resolution of the method.*

In summary, we have measured CPL spectra and used their analysis to determine the handedness of individual NCs. We have demonstrated a methodology for handedness sorting of single nano-structures based on CPL. This has been achieved for rod-shaped $Eu^{3+}$ doped (5%) $TbPO_4 \cdot D_2O$ lanthanide NCs with $10^5$ emitters illuminated with a diffraction limited spot. The low photon count needed for CPL measurement of single particles suggests that this method could be extended to single chiral molecules or molecular aggregates, provided that the emission dissymmetry is larger than $10^{-3}$, while emitting $10^7$ photons before bleaching. Considering the recent progress in the field of chiral organic chromophores[26] this challenge seems surmountable. This system could also be beneficial for bio-labeling using lanthanide-based luminescent markers, providing an additional degree of freedom besides color separation. Finally, the ability to identify and count handedness over a large number of particles together with the two dimensional spatial information forms a valuable tool to study symmetry breaking mechanisms and their kinetics.

## Methods

**Chiral lanthanide nanorod synthesis.** Terbium chloride hexahydrate (99.9%) and europium chloride hexahydrate (99.9%) (Fisher Scientific). Sodium dibasic phosphate (≥99.0%), D-tartaric acid (99%) and L-tartaric acid (≥99.5%) (Sigma). Hydrogen chloride (32%) (BioLab). Deuterated water ($D_2O$) (ARMAR Chemicals). All chemicals were used without further purification. All precursor chemicals were diluted in $D_2O$ to 100 mM stock, except the hydrogen chloride solution which was diluted in $D_2O$ to 1M. The synthesis consisted of two steps: 1) Synthesizing spherical 'seed' nanoparticles of ~5 nm diameter, and 2) Seeded growth of these seeds into larger (~1 uM on 50 nm) nanorods.

All of the syntheses were performed in $D_2O$, which enhanced the lanthanide NC luminescence, as water is a known $Eu^{3+}$ quencher[18,20,22].

1. Seed synthesis:

Solution **A** preparation: in a 20 ml scintillation vial, to 10 mL $D_2O$ were added under constant stirring, at the order as listed: 285 μL of $TbCl_3 \cdot 6H_2O$, 15 μL of $EuCl_3 \cdot 6H_2O$, 150 μL of HCl 1M and 300 μL of L- or D-TA, depending on which NC enantiomer is desired.. Solution **B** preparation: in a glass vial, under stirring, to 1 mL of $D_2O$ we added 0.6 mL of $Na_2HPO_4$ and 150 μL of HCl 1M. The solutions were allowed 30 minutes to equilibrate at the desired temperature, following which solution **B** was rapidly added to solution **A**. The addition of the phosphate solution to the $Ln^{3+}$ solution prompted the immediate precipitation of $Eu^{3+}$ doped $TbPO_4 \cdot D_2O$ NC seeds. The solution remained clear. The seeds were purified by adding 1:1 volume of acetone and centrifugation at 1070 RCF for 5 minutes, re-dissolution of the precipitate in $D_2O$, vigorous vortex and sonication for at least 20 minutes

2. Nanorod synthesis:

Solution **A** was prepared in the same manner as the seeded synthesis, only heated on a hot plate set so the solution temperature was 75°C. Solution **B** was prepared in the same manner, only 10ml of $D_2O$ was used instead of 1 ml, and it was also heated on the same hot plate as solution **A**.

The solutions were allowed 30 minutes to equilibrate at the desired temperature, following which 90 µl of the cleaned seed solution was added to solution A and immediately after solution **B** was rapidly added to solution **A**. The addition of the phosphate solution to the $Ln^{3+}$ solution prompted the precipitation of $Eu^{3+}$ doped $TbPO_4·D_2O$ NC rods, which caused the solution to grow turbid. The solution was left under heating and stirring for ~5 hours.

**Sample preparation**: The nanorods were purified in the same manner the 'seed' nanocrystals were purified. Several µl of the cleaned NR solution was drop-cast on a Formvar/Carbon 300-mesh Cu TEM grids, which was held in place on a fused-silica glass. For the racemic sample, TEM grids were made hydrophilic by $O_2$ plasma in a Diener Pico plasma oven set to 60 W for 1 minute prior to the sample deposition. The racemic NC solution was prepared by titrating one NC solution with the other NC enantiomer until the ensemble CPL of the solution was below noise level for these systems ($g_{lum}$ ~ $10^{-5}$).

**Experimental setup.**

**Single particle measurements**: The output of a supercontinuum laser (SC400 4W, Fianium) was filtered by an acousto-optic tunable filter (Gooch & Housego, AOTF 2885-02) to match the 488 nm excitation wavelength of the NCs. The output was then further spectrally cleaned by a 500 nm low pass filter (FESH0500, Thorlabs) and a 488 nm laser line (FF01-488/10-25, Semrock). A quarter waveplate (AQWP05M-600, Thorlabs) converts the linearly polarized light to circularly polarized light to avoid orientation sensitivity in the excitation of the needle-like NCs. The beam is then split by a 50:50 non polarizing beamsplitter into an excitation and reference beams. The excitation beam was then focused on the sample using a 0.85NA polarization objective (Plan NEOFLUAR 0.85 POL, Zeiss) and collected using another 0.8NA polarization objective (TU plan 0.8 POL, Nikon). The reflected light was focused onto a confocal pinhole and collected by photodiode (Two Color Sandwich Detectors, OSI Optoelectronics). In the transmitted light path a 550 nm high pass emission filter (FELH0550, Thorlabs) blocked the excitation beam and allowed the measurement of the NC's luminescence. The emitted light is then passed through a super achromatic $\lambda/4$ waveplate (APSAW-7, Astropribor) followed by a broadband wire grid linear polarizer (WP12L-UB, Thorlabs) at 45° to the fast axis of the $\lambda/4$ waveplate and fed to a PD detector. This enables measurement of circular polarized emission of a specific handedness at the full emission bandwidth. The emerging linearly polarized luminescence was then focused on a 400 µm fiber and fed into a spectrograph (Shamrock 303i, Andor) equipped with an EM-CCD camera (Newton 971, Andor) cooled to -80ºC. Each NC was measured using EM gain set to max, in full vertical binning mode. The signal was accumulated for 100-300 exposures for a total of 35-105 seconds. The TEM grid containing the NC sample was placed on a low fluorescence highly pure fused silica 170 µm cover glass (Schott HPFS 7980, Mark Optics). The cover slide was cleaned by sonication in acetone for 15 minutes, followed by sonication for 15 minutes in isopropanol, and lastly heat treated in a furnace at 1000º C for 5 hours. The final step reduced the glass residual internal fluorescence by 50%.

**Linear SVM classification:** four spectral features (see Fig. 4) were selected as input to Matlab's SVM machine learning algorithm. Each input feature was rescaled to the range of [0, 1] using $F^i_{scaled} = \frac{F^i - \min(F^i_{total})}{\max(F^i_{total}) - \min(F^i_{total})}$, where $F^i_{total}$ are all instances of the i'th feature. The training data set was divided into ten folds so that each will be used separately to train the classifier while the rest are used to as test data, thus reducing the likelihood of over fitting the model. The training and test data produced an error rate of ~0.8% based on our initial ansatz that the training samples were enantiopure. The single 'wrongly' classified NC spectrum was then rechecked visually, leading to the conclusion that the misclassified 'L' NC was in fact a 'D' handed particle. The exclusion of the 'wrong' particle brings the accuracy of the SVM classifier to 100%. Following the training, data of unknown NCs from the racemic mixture was fed into the SVM to predict their handedness.

**Solution (bulk) measurements:** measurements were performed in a home-built CPL system, using a photoelastic modulator and lock-in detection, built according to the general setup described in ref [15]. Samples were excited by a high-power near-UV Hamamatsu LED light source (7 W, 365 nm) on the sample. Light emitted from the samples was collected at 90° to the excitation, with both excitation and emission edgepass filters (FF01-424/SP-25, Semrock and FEL0450, Thorlabs) to minimize light leakage from the excitation source, passed through the monochromator and detected by a photomultiplier.

# References

1. Blackmond, D. G. The origin of biological homochirality. *Cold Spring Harb. Perspect. Biol.* **2,** a002147 (2010).

2. Wang, L. Y. *et al.* Circular Differential Scattering of Single Chiral Self-Assembled Gold Nanorod Dimers. *ACS Photonics* **2,** 1602–1610 (2015).

3. Banzer, P., Woźniak, P., Mick, U., De Leon, I. & Boyd, R. W. Chiral optical response of planar and symmetric nanotrimers enabled by heteromaterial selection. *Nat. Commun.* **7,** 13117 (2016).

4. Narushima, T., Hashiyada, S. & Okamoto, H. Optical Activity Governed by Local Chiral Structures in Two-Dimensional Curved Metallic Nanostructures. in *Chirality* **28,** 540–544 (2016).

5. Schnoering, G. *et al.* Three-Dimensional Enantiomeric Recognition of Optically Trapped Single Chiral Nanoparticles. *Phys. Rev. Lett.* **121,** 023902 (2018).

6. Vinegrad, E. *et al.* Circular Dichroism of Single Particles. *ACS Photonics* **5,** 2151–2159 (2018).

7. Karst, J., Strohfeldt, N., Schäferling, M., Giessen, H. & Hentschel, M. Single Plasmonic Oligomer Chiral Spectroscopy. *Adv. Opt. Mater.* **6,** 1800087 (2018).

8. Hassey, R., Swain, E. J., Hammer, N. I., Venkataraman, D. & Barnes, M. D. Probing the Chiroptical Response of a Single Molecule. *Science.* **314,** 1437–1439 (2006).

9. Tang, Y., Cook, T. A. & Cohen, A. E. Limits on Fluorescence Detected Circular Dichroism of Single Helicene Molecules. *J. Phys. Chem. A* **113,** 6213–6216 (2009).

10. Steinwart, I. & Christmann, A. *Support vector machines*. (Springer, 2008).

11. Riehl, J. P. & Richardson, F. S. Circularly Polarized Luminescence Spectroscopy. *Chem. Rev.* **86,** 1–16 (1986).

12. Longhi, G., Castiglioni, E., Koshoubu, J., Mazzeo, G. & Abbate, S. Circularly Polarized Luminescence: A Review of Experimental and Theoretical Aspects. *Chirality* **28,** 696–707 (2016).

13. Tohgha, U. *et al.* Ligand induced circular dichroism and circularly polarized luminescence in cdse quantum dots. *ACS Nano* **7,** 11094–11102 (2013).

14. Naito, M., Iwahori, K., Miura, A., Yamane, M. & Yamashita, I. Circularly polarized luminescent CDs quantum dots prepared in a protein nanocage. *Angew. Chemie - Int. Ed.* **49,** 7006–7009 (2010).

15. Zinna, F. & Di Bari, L. Lanthanide circularly polarized luminescence:

Bases and applications. *Chirality* **27,** 1–13 (2015).

16. Carr, R., Evans, N. H. & Parker, D. Lanthanide complexes as chiral probes exploiting circularly polarized luminescence. *Chem. Soc. Rev.* **41,** 7673–86 (2012).

17. Hananel, U., Ben-Moshe, A., Diamant, H. & Markovich, G. Spontaneous and Directed Symmetry Breaking in Chiral Nanocrystals. *Submitted* (2018).

18. Luwang, M. N., Ningthoujam, R. S. & Srivastava, S. K. Effects of Ce3+ Codoping and Annealing on Phase Transformation and Luminescence of Eu3+-Doped YPO4 Nanorods: D2O Solvent Effect. *J. Am. Chem. Soc.* **56,** 2979–2982 (2009).

19. Binnemans, K. Interpretation of europium(III) spectra. *Coord. Chem. Rev.* **295,** 1–45 (2015).

20. Bunzli, J. G. & Eliseeva, S. V. *Basics of Lanthanide Photophysics*. *Springer Series on Fluorescence* (2011). doi:10.1007/4243

21. Di, W. *et al.* Photoluminescence, cytotoxicity and in vitro imaging of hexagonal terbium phosphate nanoparticles doped with europium. *Nanoscale* **3,** 1263–9 (2011).

22. Beeby, A. *et al.* Non-radiative deactivation of the excited states of europium, terbium and ytterbium complexes by proximate energy-

matched OH, NH and CH oscillators: an improved luminescence method for establishing solution hydration states. *J. Chem. Soc. Perkin Trans. 2* **2,** 493–504 (1999).

23. Gargas, D. J. *et al.* Engineering bright sub-10-nm upconverting nanocrystals for single-molecule imaging. *Nat. Nanotechnol.* **9,** 300–305 (2014).

24. Gupta, S. K., Reghukumar, C. & Kadam, R. M. Eu3+local site analysis and emission characteristics of novel Nd2Zr2O7:Eu phosphor: Insight into the effect of europium concentration on its photoluminescence properties. *RSC Adv.* **6,** 53614–53624 (2016).

25. Viedma, C., McBride, J. M., Kahr, B. & Cintas, P. Enantiomer-specific oriented attachment: Formation of macroscopic homochiral crystal aggregates from a racemic system. *Angew. Chemie - Int. Ed.* **52,** 10545–10548 (2013).

26. Kumar, J., Nakashima, T. & Kawai, T. Circularly Polarized Luminescence in Chiral Molecules and Supramolecular Assemblies. *J. Phys. Chem. Lett.* **6,** 3445–3452 (2015).

## Contributions

E.V. and O.C. conceived and designed the experiment. E.V. built the experimental setup, wrote the control software, conducted the experiments, analyzed the data and wrote the paper. U.H. and G.M conceived and synthesized the chiral lanthanide nanocrystals. U.H. conducted ensemble experiments and performed the TEM microscopy and wrote the paper. All authors contributed to the discussion and interpretation of the results and edited the manuscript.

## Competing interests

The authors declare no competing interests.

## Corresponding author

Correspondence to Ori Cheshnovsky.